\documentclass[onecolumn,11pt]{elsarticle}
\usepackage[top=1in, bottom=1in, left=1in, right=1in]{geometry}
\usepackage[title]{appendix}
\usepackage{amsmath,amsfonts,amscd,amssymb}
\usepackage{graphicx}
\usepackage{epstopdf}
\usepackage{overpic}
\usepackage{cancel}
\usepackage{rotating}
\usepackage{url}
\usepackage{caption}
\usepackage{color}
\usepackage{rotating}
\usepackage{multirow}
\usepackage{wrapfig}
\usepackage{mathtools}
\usepackage{subeqnarray}
\usepackage{setspace}
\usepackage{standalone}
\usepackage{tikz}
\usepackage{palatino} 
\biboptions{sort&compress}
\usepackage[bottom,flushmargin,hang,multiple]{footmisc}
\usepackage{lipsum}

\definecolor{header1}{cmyk}{0,0,0,1}

\DeclareMathOperator*{\argmin}{arg\rm{}min}

\def\iu{\ensuremath{\mathrm{i}}}
\def\du{\ensuremath{\mathrm{d}}}

\begin{document}

\begin{frontmatter}
\title{\vspace{.1in}Data-driven stochastic modeling of coarse-grained dynamics with finite-size effects using Langevin regression}

\date{\today}

\author[1]{Jordan Snyder\corref{cor1}}
\ead{jsnyd@uw.edu}
\author[2]{Jared L. Callaham}
\author[2]{Steven L. Brunton}
\author[1]{J. Nathan Kutz}
\address[1]{Department of Applied Mathematics, University of Washington}
\address[2]{Department of Mechanical Engineering, University of Washington}
\cortext[cor1]{Corresponding Author}

\begin{abstract}
    Obtaining coarse-grained models that accurately incorporate finite-size effects is an important open challenge in the study of complex, multi-scale systems. We apply Langevin regression, a recently developed method for finding stochastic differential equation (SDE) descriptions of realistically-sampled time series data, to understand finite-size effects in the Kuramoto model of coupled oscillators. We find that across the entire bifurcation diagram, the dynamics of the Kuramoto order parameter are statistically consistent with an SDE whose drift term has the form predicted by the Ott-Antonsen ansatz in the $N\to \infty$ limit. We find that the diffusion term is nearly independent of the bifurcation parameter, and has a magnitude decaying as $N^{-1/2}$, consistent with the central limit theorem. This shows that the diverging fluctuations of the order parameter near the critical point are driven by a bifurcation in the underlying drift term, rather than increased stochastic forcing.\\
\end{abstract}
\end{frontmatter}

\section{Introduction}
Collective behavior is an important and fascinating set of phenomena that occurs in a wide array of both naturally occurring and engineered systems.
Examples include flocking and collective foraging of groups of animals, synchronous flashing of fireflies, correlated activity of neurons in sensory systems, and rolling blackouts in electric power grids.
Characterizing such phenomena relies on a systematic construction of coarse-grained, reduced-order models that parsimoniously capture the physics of system-wide (macro-scale) observables.
For instance, the Kuramoto model of coupled oscillators, with its synchronization phase transition, is a paradigmatic model of collective behavior, with the construction of reduced-order models having been successfully developed over the many years since its introduction.
In addition, much is known about the statistics of fluctuations near the phase transition for finite system sizes, and their scaling with system size has been shown to be anomalous relative to typical statistical physics problems.
However, comparatively little work has been done to systematically incorporate finite-size effects in reduced-order models of coupled oscillators.
In this paper, we apply Langevin regression, a recently introduced method for extracting SDE descriptions of realistic time-series data~\cite{Callaham2020}, to the problem of obtaining physically interpretable stochastic models of finite-size effects in the Kuramoto model.

The determination of collective behavior is related to the construction of closure models which aim to characterize 
the effect of unresolved degrees of freedom upon observed coarse-grained dynamics.  The unresolved and unmeasured degrees of freedom lead to the manifestation of randomness and memory effects, even when the underlying laws of motion are deterministic and memoryless~\cite{Zwanzig1961,Zwanzig1973,Lin2021}.  Thus it is natural to seek a stochastic differential equation (SDE) as a coarse-grained model. We may further desire that our coarse-grained model be interpretable, suggesting that it may be appropriate to regress to a given family of SDEs.

Coarse-grained model reductions are often considered with the canonical Kuramoto model of coupled oscillators~\cite{kuramoto1975self}, which has long been studied as an example of collective behavior~\cite{aschoff1981circadian,Strogatz1989,Frank2000,Arenas2006,Li2008}. The Kuramoto model features a composition of many coupled, nonlinear oscillatory units with distinct natural frequencies, where the pairwise coupling tends to drive phases to synchronicity. Importantly, the Kuramoto
model provides tractable coarse-grained approximations in certain special limits~\cite{strogatz2000kuramoto,ott2008low} despite its rich behavioral phenomenology~\cite{Panaggio2015,Bick2018,Zhang2019a}. In addition, the Kuramoto model and its variants describe a number of synchronization phenomena in domains of diverse nature such as coupled Josephson junctions~\cite{wiesenfeld1998frequency}, neuroscience~\cite{varela2001brainweb}, chemical oscillators~\cite{kiss2002emerging,zlotnik2016phase}, and the power grid~\cite{dorfler2013synchronization}.
The synchronization observed in the Kuramoto model allows for asymptotic approximations of a closure model.  In the seminal work of Ott and Antonsen~\cite{ott2008low}, they showed that in the $N\to\infty$ limit and under certain conditions on the distribution of natural frequencies and initial phases, the center of mass of a population of Kuramoto oscillators remarkably obeys an autonomous ODE which undergoes a pitchfork bifurcation as the coupling strength is varied, i.e. an {\em exact} closure model can be constructed.   Other studies have focused on complex coupling topologies, proposing techniques using spectral information to merge nodes together~\cite{Gfeller2008} or otherwise systematically discarding irrelevant degrees of freedom~\cite{Izumida2013}. Still others take a more strictly data-driven approach and seek closed equations of motion for low-order moments of the distribution of phases~\cite{Moon2006,Rajendran2011} or to identify good coarse-grained variables via manifold learning techniques~\cite{Thiem2020}. Finally there are approaches that employ a ``collective coordinate'' ansatz governing the phase of each oscillator within a phase-locked cluster, and thereby arrive at a closed equation of motion~\cite{Gottwald2015,Hancock2018,Smith2019,Smith2020,Yue2020}.

Our aim is to construct an order-parameter description for finite-sized Kuramoto systems, and more broadly provide mathematical methods for characterizing finite-sized coarse-grained models.  This has been studied previously using nonequilibrium statistical mechanics~\cite{Buice2007}, whose results are indirect and difficult to interpret.  Others have considered the 
ensemble-variance in the steady state order parameter magnitude, and dynamical fluctuations under quenched disorder~\cite{Daido1987,Daido1988,Daido1989,Daido1990,Daido2015,Hong2015,Hong2017}. While precise finite-size scaling results can be obtained for ensemble statistics of the steady-state order parameter, investigations of dynamical fluctuations were limited to numerical simulation. Most importantly, investigations of dynamical fluctuations did not take the step of interpreting the order parameter dynamics as governed by an SDE.  In this paper, we build on prior work concerning coarse-grained modeling of Kuramoto oscillators by applying Langevin regression to simulated trajectories. We obtain physically meaningful SDE models for finite-size effects, demonstrating that the exotic finite-size scaling behavior near the critical point is consistent with white noise forcing with a magnitude that scales as $N^{-1/2}$. We conclude that critical finite-size scaling behavior is driven mainly by instabilities in the underlying dynamics rather than anomalously large forcing.

\begin{figure}
    \centering
    \includegraphics[width = \textwidth]{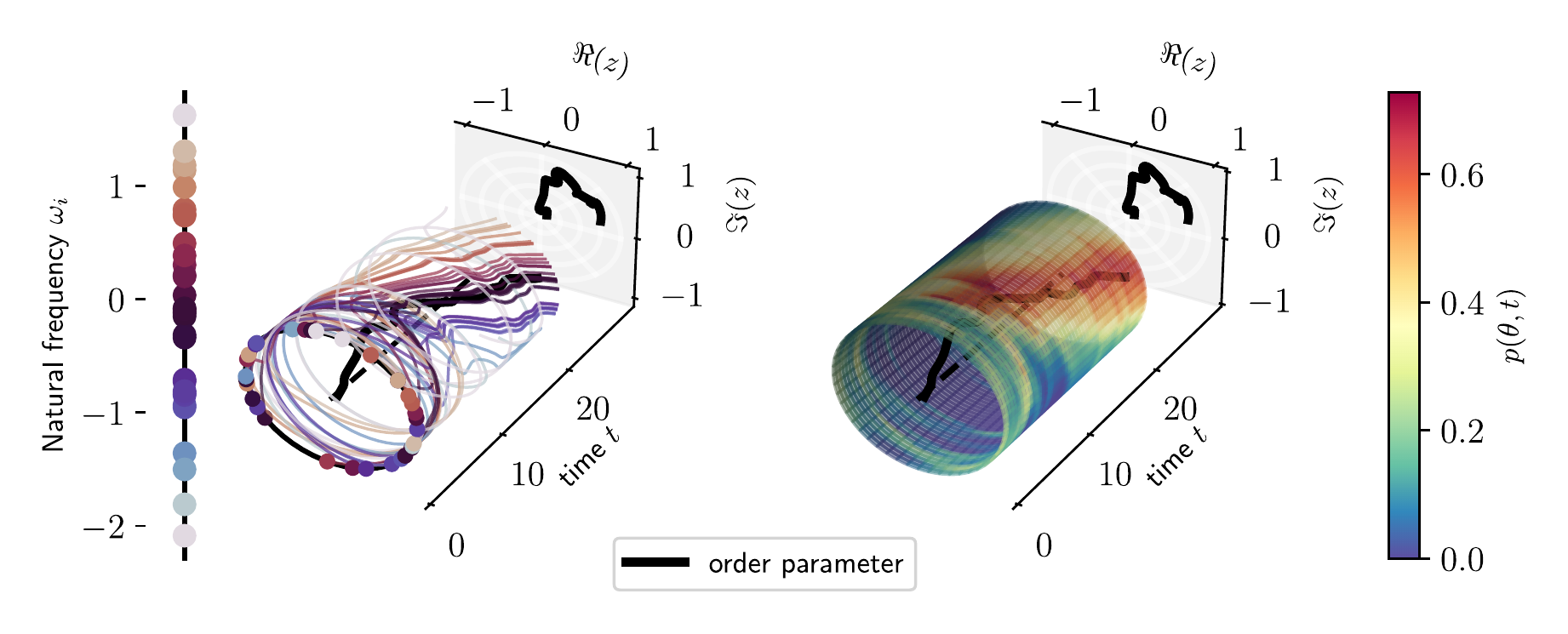}
    \caption{Emergence of synchronization in a system of $N=32$ coupled Kuramoto oscillators, and the associated order parameter, at the level of individual phases (left) and probability density (right). On the left, each colored line is the phase of one oscillator through time, plotted on a circle in the plane perpendicular to the $t$-axis. Line color corresponds to the oscillator's natural frequency $\omega_i$. On the right, color indicates the number density of oscillators as a function of phase $\theta$ and time $t$. In both plots, the solid black curve depicts the order parameter $z = \langle \exp(\iu \theta_j)\rangle_{j=1}^{N}$ over time, and the dashed line is the origin. The trajectory of the order parameter is also projected onto the back plane. Initially, the oscillators' phases are spread evenly around the circle and $|z|\sim 0$. Due to the attractive coupling, the oscillators' phases quickly approach one another, forming a synchronized cluster of oscillators that precesses with a single frequency. The magnitude of the order parameter quantifies how synchronized the system is as a whole, while its phase tracks the phase of the dominant synchronized cluster. Note that the oscillators participating in the synchronized cluster are those with natural frequencies near the middle of the distribution, while oscillators with large (positive or negative) natural frequencies continue to drift.}
    \label{fig:trajectories-on-a-tube}
\end{figure}

\section{Background}
Networks of coupled oscillators are an extremely popular object of study for applied mathematicians interested in collective behavior, owing to their rich phenomenology, relative tractability, and broad abstract representation of dynamic behavior. A paradigmatic model of coupled oscillators is the Kuramoto model:
\begin{equation}
\dot{\theta_i} = \omega_i + {1 \over N} \sum_{j=1}^{N} K_{ij}\sin(\theta_j - \theta_i), \hspace{1cm} i = 1, 2, \dots, N
\label{eq:kuramoto}
\end{equation}
where $\theta_i$ is the \emph{phase} of oscillator $i$, $\omega_i$ is its \emph{natural frequency}, and $(K_{ij})$ is the \emph{coupling matrix}. A fundamental result due to Kuramoto~\cite{kuramoto1975self} is that if coupling is mean-field (i.e. $K_{ij}\equiv K$ for all $i,j$) and natural frequencies are distributed according to a symmetric, unimodal probability density, then there is a critical coupling strength $K_c$ such that for $K>K_c$ the system is (partially) synchronized while for $K<K_c$ the system is incoherent. In particular, synchrony can be measured by the \emph{order parameter} $z\coloneqq \langle \exp(\iu \theta_j)\rangle_{j=1}^{N}\in \mathbb{C}$, whose magnitude $|z|$ is near zero if oscillators' phases are spread evenly around the unit circle, and is nonzero if oscillators' phases break rotational symmetry and cluster around a preferred phase. See Fig.~\ref{fig:trajectories-on-a-tube} for a visualization.

Next, we review two major threads of research concerning the Kuramoto model. First is research that seeks equations of motion for reduced or coarse-grained degrees of freedom, towards obtaining a mechanistic understanding of collective behavior, and is generally carried out in the thermodynamic ($N\to \infty$) limit. Second is research that seeks to understand synchronization as a phase transition by investigating how the properties of finite systems behave as $N\to\infty$. Our work bridges these threads by learning stochastic reduced-order models for finite systems and studying their scaling with $N$.

\subsection{Dimension Reduction Methods}

There has been extensive work studying the conditions under which the Kuramoto model admits description as a lower-dimensional dynamical system~\cite{strogatz2000kuramoto, Arenas2008, Rodrigues2016, Bick2019}.  In the case that all natural frequencies are the equal, it is known that for any $N\ge 3$, the full $N$-dimensional phase space is foliated by invariant two-dimensional tori~\cite{Watanabe1993}, and the dynamics on those tori are generated by the action of the M\"{o}bius group~\cite{Marvel2009}. In effect, this means that the dynamics of the Kuramoto model with identical natural frequencies is two-dimensional, regardless of the number of oscillators $N$.

Concerning different natural frequencies, Ott and Antonsen~\cite{ott2008low} showed that if coupling is through a mean-field (i.e. $K_{ij} \equiv K$ for all $i, j$) and if natural frequencies $\omega_i$ are drawn from a Cauchy distribution with location $\Omega$ and scale $\delta$, then there is a closure for the order parameter ${z \coloneqq \langle \exp(\iu \theta_j)\rangle_{j=1}^{N}\in \mathbb{C}}$, namely
\begin{equation}
\dot{z} = \iu(\Omega + \iu \delta) z + {K \over 2}(z - |z|^2 z).
\label{eq:OA_ansatz_base_case}
\end{equation}
This equation is valid in the $N\to \infty$ limit and technically only for certain (distributional) initial conditions; a follow-on paper~\cite{Ott2009} established under mild assumptions that the manifold in question is attractive and so the above equation should hold after some initial transient, while recent work has shown that the Ott-Antonsen (OA) manifold is not attracting for finite $N$~\cite{Engelbrecht2020}. Equivalent results hold for the case that the oscillators are organized into modules, where within each module the oscillators' natural frequencies are drawn from a Cauchy distribution and the coupling strength between oscillators $i$ and $j$ depends only on the modules they belong to. If the distribution of natural frequencies is not Cauchy but a different rational function of some order, then a similar reduction is possible, but with one variable for each pair of poles of the natural frequency distribution~\cite{Skardal2018a}.

Another analytical tool for deriving reduced models of coupled oscillator systems is the ``collective coordinate'' method introduced by Gottwald~\cite{Gottwald2015}. This method represents the state of the population of oscillators by a shape function together with a time-varying constant of proportionality, the collective coordinate, such that the residual phase of any phase-locked oscillator is given by the collective coordinate multiplied by the shape function evaluated at its natural frequency. The resulting equations offer remarkably accurate predictions for finite networks, including the scaling behavior of the position of the critical point~\cite{Gottwald2015}, have been applied to more complex network structures~\cite{Smith2020,Smith2019}, and have been used as inductive biases to guide data-driven discovery of coarse-grained models~\cite{Snyder2020a}.

Finally, there are data-driven approaches to discovering coarse-grained models for coupled oscillator systems. Prominent among these is the equation-free (EF) framework, which estimates the time derivative of a coarse variable based on projections of numerically integrated fine-grained dynamics. The EF approach has been used to obtain accurate coarse-grained numerical models of oscillator systems with a collective coordinate-type approach~\cite{Moon2006} and for networks with spectral gaps~\cite{Rajendran2011}. More recently, this approach has been combined with manifold learning techniques to learn the coarse-graining map itself in addition to the dynamics of the coarse variables~\cite{Thiem2020}.

\subsection{Finite-Size Scaling}

While reduced-order modeling has proven to offer accurate and physically meaningful descriptions of populations of coupled oscillators, little of this work deals directly with the manner in which models for finite systems converge as $N\to\infty$. On the other hand, much is known about the finite-size scaling of certain summary statistics in the Kuramoto model from a statistical physics standpoint.

The synchronization transition in the Kuramoto model is characterized by a continuous yet abrupt change in synchrony and a diverging susceptibility, or sensitivity to external forcing, closely related to temporal variance of the order parameter~\cite{Daido1988,Daido1989,Daido1987,Daido1990}. Both the abruptness of the change in synchrony and the divergence of susceptibility are smoothed out in finite systems, and that smoothing is described by finite-size scaling relations. Progress on this problem for the Kuramoto model has focused on applying scaling ans\"{a}tze to the self-consistent equation that constrains the stationary value order parameter, as well as to averages and variances of the order parameter computed via numerical simulation~\cite{Hong2015,Hong2017,Coletta2017}.

While fundamental, the results described above still leave questions unanswered regarding the origin of the particular scaling relations that are observed. Through stochastic modeling over a range of system sizes, we demonstrate that the scaling relations present in the Kuramoto model are consistent with an SDE whose drift term has the form predicted by the $N\to\infty$ theory due to Ott-Antonsen, and whose diffusion term, capturing finite-size fluctuations, scales as $N^{-1/2}$, consistent with the central limit theorem.

\section{Methods}

\subsection{Model System and Analytics}
\label{sec:background-langevin}
\begin{figure}
    \centering
    \begin{overpic}[width=\textwidth]{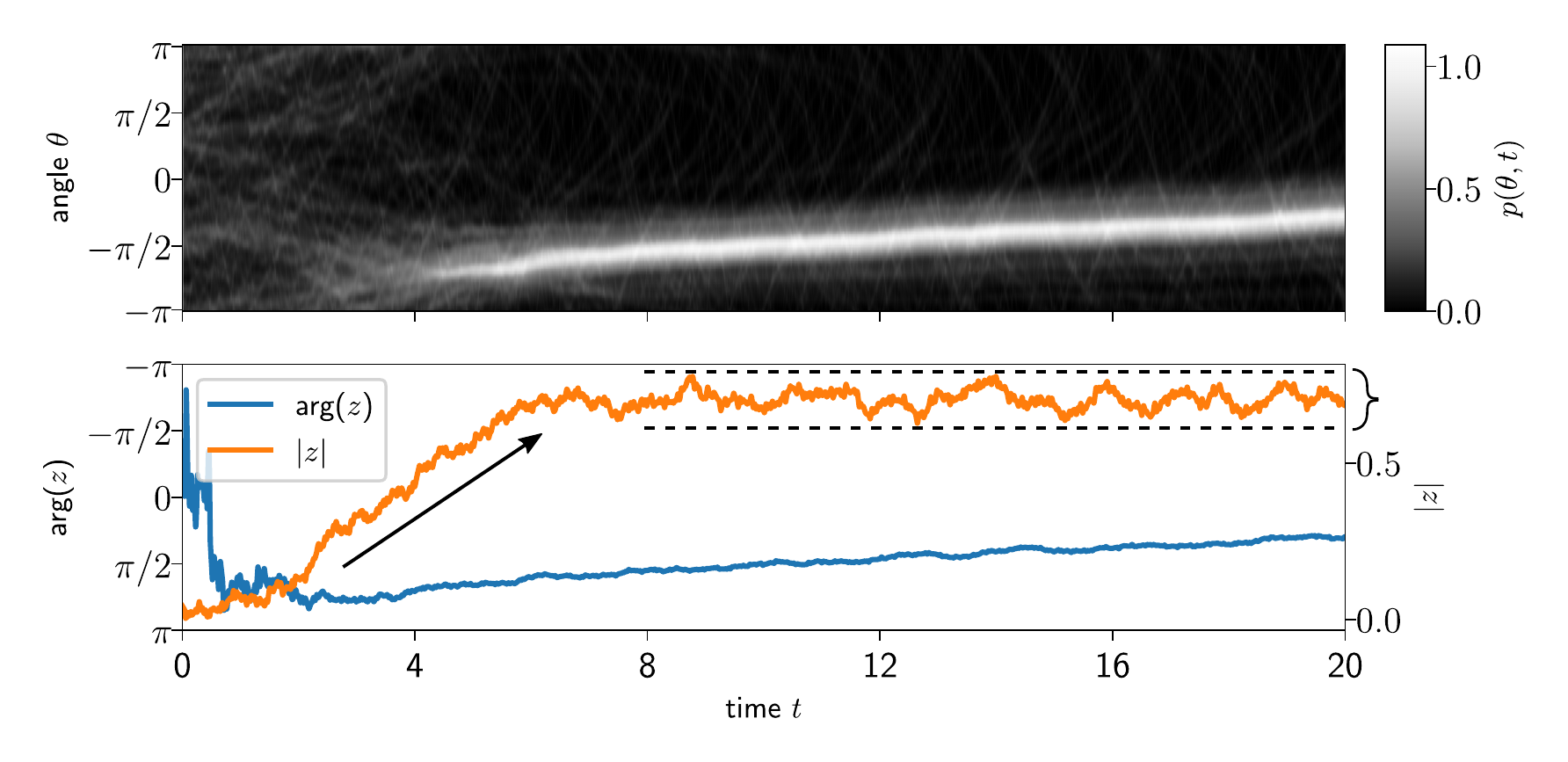}
        \put(30, 15){increasing synchrony}
        \put(88, 22.5){ \parbox{0.7in}{finite-size effects}}
    \end{overpic}
    \caption{Visualization of the emergency of synchronization in a system of $N=200$ Kuramoto oscillators in terms of the probability density (top) and order parameter (bottom). The density plot shows that the oscillators' phases rapidly align with one another, forming an obvious peak in density starting at time $t\gtrsim 5$. The degree of alignment can be quantified by the magnitude of the order parameter $z$, depicted in orange in the bottom plot. Rather than settling down to a constant value, the magnitude of the order parameter fluctuates due to the system's finite size. The angle of the order parameter meanwhile closely tracks the location of the peak in oscillator density. Oscillators' natural frequencies were drawn from a Cauchy distribution with width $\delta=1/2$ and coupled with coupling strength $K=2$.}
    \label{fig:density-evolution-pde}
\end{figure}

We now compute the form of the SDE that we will fit to the dynamics of the order parameter $z$. According to the Ott-Antonsen ansatz, $z$ should evolve according to an ODE of the form
\begin{equation}
    \dot{z} = \lambda z + \mu |z|^2 z
    \label{eq:OA_complex_ODE}
\end{equation}
where $\lambda = \iu\Omega - \delta + K/2$ and $\mu = -K/2$. The solution to this ODE either approaches $z=0$, or a fixed point or limit cycle with $|z|^2 = -\lambda / \mu$. However, it is clear from simulations that when $N$ is finite, the value of $z$ exhibits fluctuations that persist through time (see Fig.~\ref{fig:density-evolution-pde}). The reason for these fluctuations is that when $N$ is finite, the distribution of oscillators' phases is necessarily discrete, and so cannot approach the smooth distribution described by the OA anstaz that allows $z$ to evolve autonomously.

We therefore suppose that when $N$ is finite, the statistics of $z(t)=r(t)\exp(\iu \Theta(t))$ will be well described by an SDE whose drift term has the form of Eq.~(\ref{eq:OA_complex_ODE}), and whose noise term quantifies finite-size effects.
Namely, we suppose that for finite $N$, the order parameter is governed by an SDE of the form
\begin{equation}
    \du z = \left( \lambda z + \mu |z|^2 z \right) \du t + \sigma \exp(\iu\Theta) \du W
    \label{eq:complex_SDE_form}
\end{equation}
where $\sigma\in\mathbb{R}$ depends on $N$, and $W = W_1 + \iu W_2$ is a complex-valued Brownian motion with independent real and imaginary parts $W_1$ and $W_2$ respectively. Because $W_1$ and $W_2$ are independent, adjusting the phase of the noise term by $\exp(\iu\Theta)$ does not affect the distribution of the solution, while it does simplify the next step of the analysis. We emphasize that modeling finite-size effects as Brownian motion reflects an assumption that fluctuations have a finite correlation time, which may or may not be satisfied in practice, and that our goal is not to exactly model a given trajectory $z(t)$, but rather its statistics.

In addition to fluctuations in time, finite Kuramoto systems may also exhibit uncertainty in their parameter values. For instance, $\lambda = \iu\Omega - \delta + K/2$ depends on population statistics: the location, $\Omega$, and scale, $\delta$, of the distribution of natural frequencies. However, the realized set of natural frequencies is a finite sample, and thus may not be precisely representative of the infinite population. We therefore treat $\lambda$ and $\mu$ as unknowns to be fit to observed data, in addition to $\sigma$.

For simplicity, we recast the complex-valued SDE (\ref{eq:complex_SDE_form}) into polar coordinates $z=r\exp(\iu\Theta)$ so that
\begin{align}
    \du r &= \left(\lambda_r r + \mu_r r^3 + {\sigma^2 \over 2r}\right) \du t + \sigma\du W_1  \label{eq:radial_SDE_form}\\
    \du \Theta &= \left(\lambda_i + \mu_i r^2 \right) \du t + {\sigma\over r} \du W_2
\end{align}
where subscripts $i$ and $r$ denote real and imaginary values of the respective variables. Note the $\sigma^2/2r$ term in the drift function for $r$, which arises due to the It\^{o} formula for the change of variables~\cite{Gardiner1996} and ensures that $r$ never becomes negative despite being subject to additive noise. Since $r$ is the variable associated directly with the symmetry-breaking synchronization transition, and since the dynamics of $r$ do not depend on $\Theta$, we hereafter focus exclusively on the dynamics of $r$.

We wish to investigate whether or not the dynamics of $r$ for finite Kuramoto systems are consistent with a model of the form (\ref{eq:radial_SDE_form}), and if so, what are the properties of the coefficients $\lambda, \mu$, and $\sigma$ as a function of system size $N$ and coupling strength $K$?

\subsection{Data Generation}

To investigate these issues, we performed numerical simulations of the Kuramoto model for different values of the number of oscillators $N$, and the coupling strength $K$. For all experiments, natural frequencies were drawn i.i.d. from a Cauchy distribution with location $\Omega = 0$ and scale $\delta = 1/2$, meaning that the critical coupling strength is $K_c = 2/\pi g(0) = 1$~\cite{kuramoto1975self}.
We define the bifurcation parameter $\epsilon \coloneqq K-K_c$, such that the symmetry-breaking transition occurs at $\epsilon = 0.$ 
The number of oscillators was taken to be each of $\{64, 256, 1024, 4096\}$, $\epsilon$ was set at 20 evenly spaced values between -1 and 1.
For each value of $(N, \epsilon)$, we simulate an ensemble of 100 systems, each with a different sample of natural frequencies $\{\omega_i\}$ and initial phases $\{\theta_i(0)\}$.
For each sample we integrate the ODE~\eqref{eq:kuramoto} to a final time $T = 10^3$ with a time resolution of $\Delta t = 10^{-3}$, using a 5th-order Tsitouras scheme~\cite{Rackauckas2017differentialequations}. Finally we coarse-grained the resulting time series $\{\theta_i(t)\}$ to obtain
\begin{equation}
    r(t)\exp(\iu \Theta(t)) \coloneqq {1\over N} \sum_{j=1}^{N} \exp(\iu \theta_j(t)).
\end{equation}
Our goal is to obtain an SDE model that is statistically consistent with the observed time series $r(t)$, after discarding an initial transient. Specifically, we seek an SDE of the form (\ref{eq:radial_SDE_form}).

\subsection{Inference Procedure}

\begin{figure}
    \vspace{-.15in}
    \includegraphics[width=\textwidth]{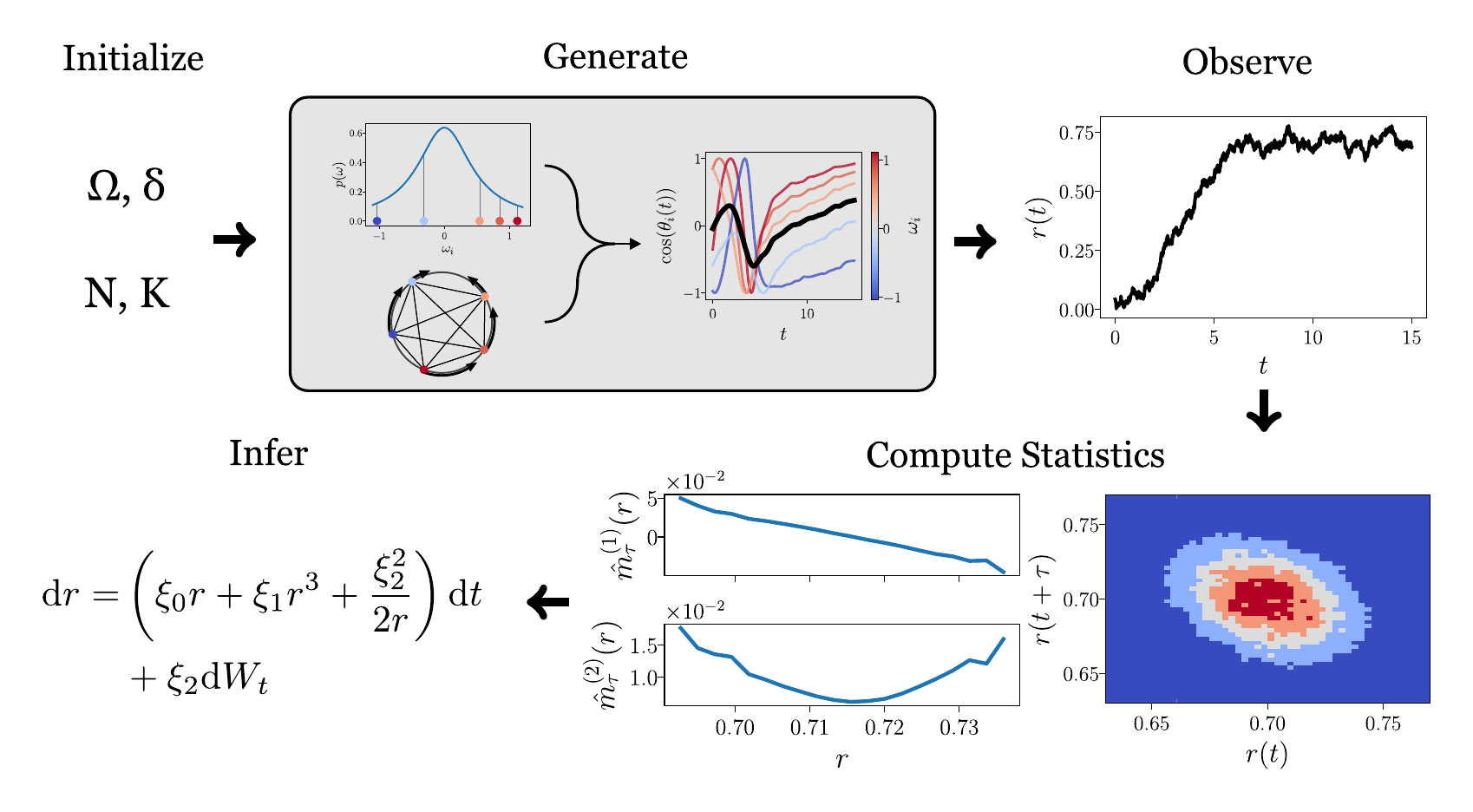}
    \vspace{-.35in}
    \caption{Schematic of our analysis pipeline. We first sample $N$ i.i.d. natural frequencies $\{\omega_i\}$ from a Cauchy distribution with location $\Omega$ and scale $\delta$, then integrate the Kuramoto ODE (\ref{eq:kuramoto}) forward in time from randomly assigned initial phases. We then coarse-grain the trajectory $\{\theta_i(t)\}$ to obtain the complex order parameter $z(t) = r(t)\exp(\iu \Theta(t))$, and compute the joint distribution $p(r(t), r(t+\tau))$. From this joint distribution we compute the first two finite-time Kramers-Moyal averages (\ref{eq:KM-coefficients}), and finally use Langevin regression to infer SDE coefficients $\xi$ that are consistent with the measured finite-time Kramers-Moyal averages.}
    \label{fig:schematic}
\end{figure}

Based on the discussion in Sec.~\ref{sec:background-langevin}, we expect that the coarse-grained SDE will be of the form
\begin{equation}
    \du r = \left(\xi_0 r + \xi_1 r^3 + {\xi_2^2 \over 2r}\right) \du t + \xi_2 \du W_t,
\end{equation}
where the parameters $\xi = (\xi_0, \xi_1, \xi_2)$ are to be identified with the recently proposed Langevin regression method for identifying stochastic differential equations from data~\cite{Callaham2020}.
This method solves both the forward and adjoint Fokker-Planck equations to enforce consistency with both the finite-time conditional moments and the steady-state probability distribution (PDF), both computed after discarding an initial transient. A schematic of the overall procedure performed in this paper is given in Fig.~\ref{fig:schematic}.

For each of the replicate experiments described above, we computed the empirical PDF $\hat{p}(r)$ and the first two ($n=1, 2$) empirical finite-time conditional moments $\hat{m}^{(n)}_\tau(r)$, defined as
\begin{align}
\label{eq:KM-coefficients}
    \hat{m}^{(n)}_\tau(r) &= \left(\mathbb{E}\left[ (r(t+\tau) - r(t))^n \vert r(t) = r \right]\right)^{{1\over n}}.
\end{align}
The power of $1/n$ ensures that the conditional moments of different orders are dimensionally consistent with one another, so that they can be appropriately weighted in the objective function that we define below.
For a stationary process near thermal equilibrium with a sampling rate much faster than any natural time scale of the system, these moments can be directly related to the drift and diffusion functions~\cite{Risken1996book,Siegert1998, Friedrich2011, Boninsegna2018}.
However, for many closure problems the neglected degrees of freedom have unresolved dynamics that lead to non-Markovian memory effects upon truncation.

Sampling the system too quickly will then invalidate the assumption of white-in-time forcing underlying the convergence of Eq.~\eqref{eq:KM-coefficients} to a useful estimator of the drift and diffusion.
On the other hand, coarse sampling rates lead to significant finite-time distortion of the conditional moments~\cite{Ragwitz2001}.
For example, the evolution of the order parameter is smooth at short times; Eq.~\eqref{eq:KM-coefficients} would significantly underestimate the drift and diffusion coefficients at fast sampling rates.
We therefore compute the conditional moments using $\tau = 0.1$.
Langevin regression uses the adjoint Fokker-Planck equation to correct for finite-time effects~\cite{Lade2009, Honisch2011}.
The empirical conditional moments $\hat{m}_\tau^{(n)}(r)$ are then compared against the finite-time conditional moments $m_\tau^{(n)}(r, \xi)$ that would be observed given a set of parameters $\xi$ (see Ref.~\cite{Callaham2020} for details). Because the computed trajectories are finite, we estimate $\hat{m}_\tau^{(n)}(r)$ by computing a histogram with equally-sized bins centered at locations $\{r_i\vert i = 1, \dots, M\}$. In the following, we use $M=40$ equally-sized bins that cover full range of observed $r$ values.

Langevin regression solves the following optimization problem:
\begin{equation}
    \xi = \argmin_\xi \sum_{n=1}^2 \sum_{i=1}^{M} w_i^{(n)} \left[m_\tau^{(n)}(r_i, \xi) - \hat{m}_\tau^{(n)}(r_i)\right]^2 + \eta_\text{KL} \mathcal{D}_\text{KL} \left(\hat{p}(r)\Vert p(r, \xi)\right) + \eta_{\ell_1} \Vert s \odot \xi \Vert_1,
\end{equation}
where $\odot$ denotes the entrywise (Hadamard) product of vectors, the weight variables $w^{(n)}_i$ reflect pointwise uncertainty in the empirical estimate of the moments, and the $\eta$ variables control the relative contributions of each of the terms in the cost function.  We have added two regularizing terms. The Kullback-Leibler (KL) divergence $\mathcal{D}_\text{KL}$ is a statistical measure of difference between probability distributions, and is defined as
\begin{equation}
    \mathcal{D}_\text{KL}(p\Vert q) = \int p(x) \log\left({p(x)\over q(x)}\right).
\end{equation}
The KL regularization ensures that the inferred model has a steady-state PDF comparable to the empirical one. Finally, we normalize the weights $w^{(n)}_i$ so that they integrate to unity. Because the domain is discrete, this means enforcing the condition
\begin{equation}
    \sum_i (r_{i+1}-r_i) w^{(n)}_i = 1, \quad n = 1, 2
\end{equation}

The final term in the optimization problem, the $\ell_1$ norm of the rescaled coefficient vector $s \odot \xi$, is new relative to the original method.
The subcritical systems ($\epsilon < 0$) are linearly stable about the origin, introducing a redundancy between the linear and cubic terms.
The $\ell_1$ regularization promotes sparse coefficient vectors, encouraging the optimization problem to select zero or near-zero coefficients where these are consistent with the data.
The rescaling vector $s$ allows us to adjust the magnitudes of the entries of $\xi$ so that they are comparable to each other. As we expect the noise coefficient $\xi_2$ to decay as $N^{-1/2}$, we use $s = (1, 1, N^{1/2})$. We used regularization parameters $\eta_\text{KL} = 10^{-3}$ and $\eta_{\ell_1} = 10^{-4}$, as we found these values to prevent both blowup and premature dropout of inferred coefficients. Optimization was carried out using the Nelder-Mead method as implemented in \texttt{scipy.optimize}~\cite{Virtanen2020}.

\section{Results}
\begin{figure}
    \centering
    \includegraphics[width =\textwidth]{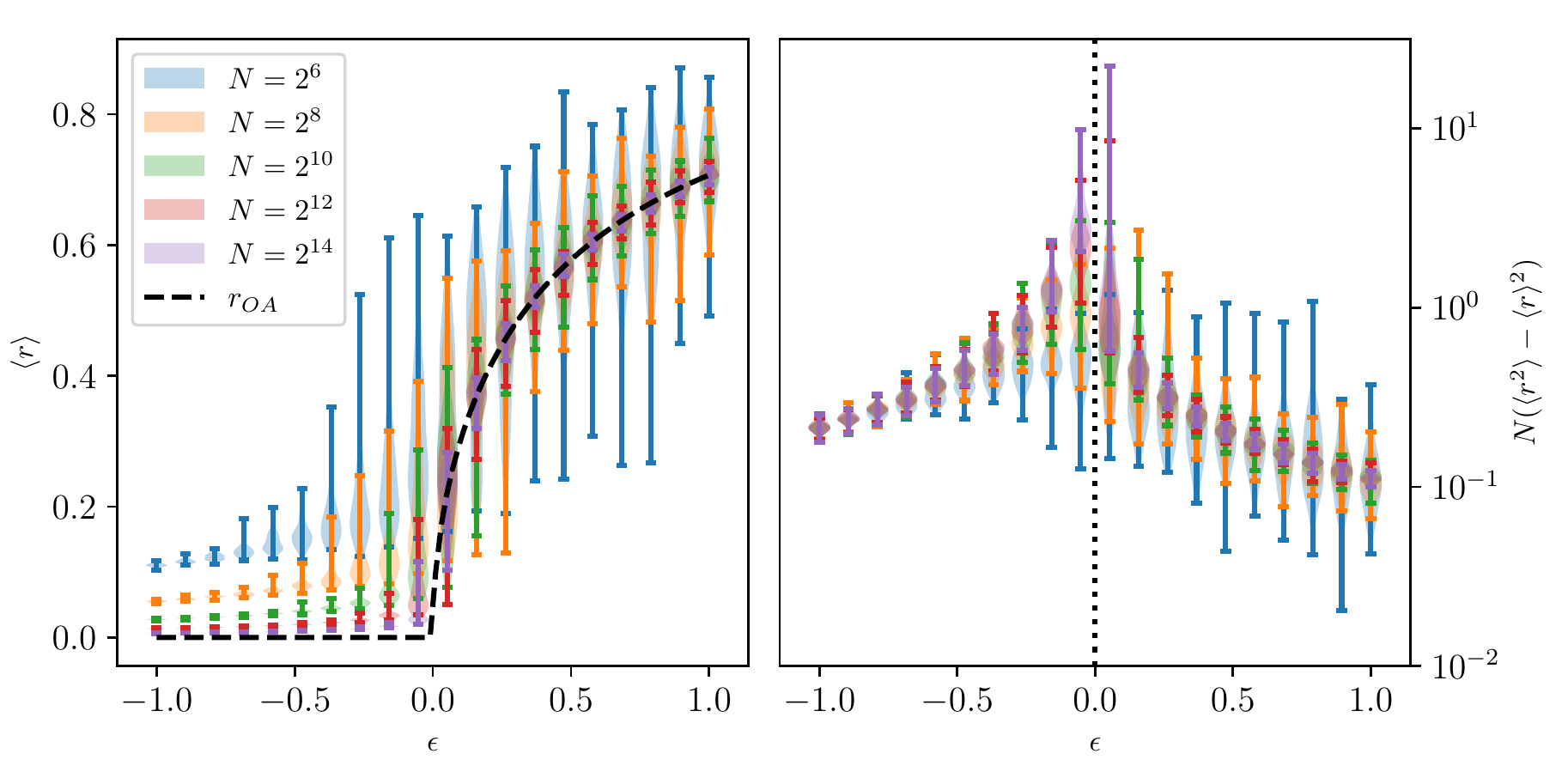}
    \caption{Left: Violin plots of the distribution of the temporal average of $r$ as a function of bifurcation parameter $\epsilon = K-K_c$ and $N$. For each value of $(N, \epsilon)$, we sample 100 Kuramoto systems and compute the time-average of $r$ (after discarding the transient), and depict the distribution of these 100 values as a vertical violin. Vertical lines with tick-ends depict the max and min of each set of 100 values. Dashed line is $r_{OA}$, the value of the fixed point of the ODE model derived from the Ott-Antonsen ansatz, Eq.~(\ref{eq:OA_ansatz_base_case}). Note that to the right of the critical point ($\epsilon>0$), values cluster around the $r_{OA}$ with a variance that shrinks with $N$. To the left of the critical point, values also approach $r_{OA}$ as $N$ increases, but are offset from zero due to the singularity at the origin (see the term $\sigma^2/2r$ in the drift term of Eq.~(\ref{eq:radial_SDE_form})).
    Right: Violin plots of the distribution of the temporal variance in $r$, scaled by $N$. As on the left, for each $(N, \epsilon)$ we observe 100 Kuramoto systems, compute the temporal variance of $r$, and depict the distribution of these 100 values as a vertical violin. The ensure comparability across $N$, we scale the variance by $N$. Compare to Hong et al., Figure 1~\cite{Hong2015}. Note that while far from the critical point ($\epsilon = 0$) variance appears to be inversely proportional to $N$ (since $\chi \coloneqq N(\langle r^2\rangle - \langle r \rangle^2)\sim \mathcal{O}(1)$), as would be predicted by the central limit theorem, near the critical point $\chi$ exhibits a peak whose height grows with $N$, meaning that the variance decays more slowly than $N^{-1}$
    }
    \label{fig:r_avg_overviews}
\end{figure}

We first report summary statistics about the trajectories under study. Fig.~\ref{fig:r_avg_overviews} (left) shows the time-average of the order parameter, relative to the value predicted by the Ott-Antonsen ansatz, $r_{OA} = \max(0,\sqrt{ \epsilon /(\epsilon + 1)})$.
To the right of the bifurcation ($\epsilon>0$), the distribution of the time-average of $r(t)$ has its peak near the Ott-Antonsen value, and has a width that shrinks with $N$. On the left, however, the presence of noise makes it impossible for $r$ to remain near zero, leading to a distribution of time-average $r(t)$ values with a nonzero peak whose location decreases with $N$.


Next, Fig.~\ref{fig:r_avg_overviews} (right) depicts the variance of $r(t)$ over time, as a function of $\epsilon$ and $N$. In particular we are visualizing the variance scaled by $N$, which we refer to as $\chi = N\langle (r-\langle r \rangle)^2 \rangle$. If at any given time $t$ the complex phases $\{\exp(\iu\theta_i(t))\}$ were independent, then we would expect their mean, $z$, to exhibit finite-sample fluctuations scaling as $N^{-1/2}$ by virtue of the central limit theorem (CLT). That is, we would expect $\chi \sim \mathcal{O}(1)$, and this appears to be the case away from the critical point. Near the critical point, however, substantial correlations build up among the individual phases, and CLT scaling is no longer a good approximation.

Based on extensive numerical study, Hong et al. report that the peak $\chi_\text{max}$ of $\chi$ with respect to $\epsilon$ scales as $\chi_\text{max}(N) \sim N^{0.40}$. Finally, Fig. 1 of~\cite{Hong2015} reports the ensemble average of $\chi$ over 100-1000 trials, obtaining highly precise estimates of the population average of $\chi$; we remark that the full distribution of $\chi$ has nontrivial scaling properties with $N$. Namely, in the far subcritical regime ($\epsilon<0$) the distribution appears to be independent of $N$, while in the supercritical regime ($\epsilon>0$) the distribution of $\chi$ appears to narrow with increasing $N$. We explore explanations for this fact in Appendix~\ref{sec:chi-scaling-appendix}.

\begin{figure}
    \centering
    \vspace{-.25in}
    \includegraphics[width=\textwidth]{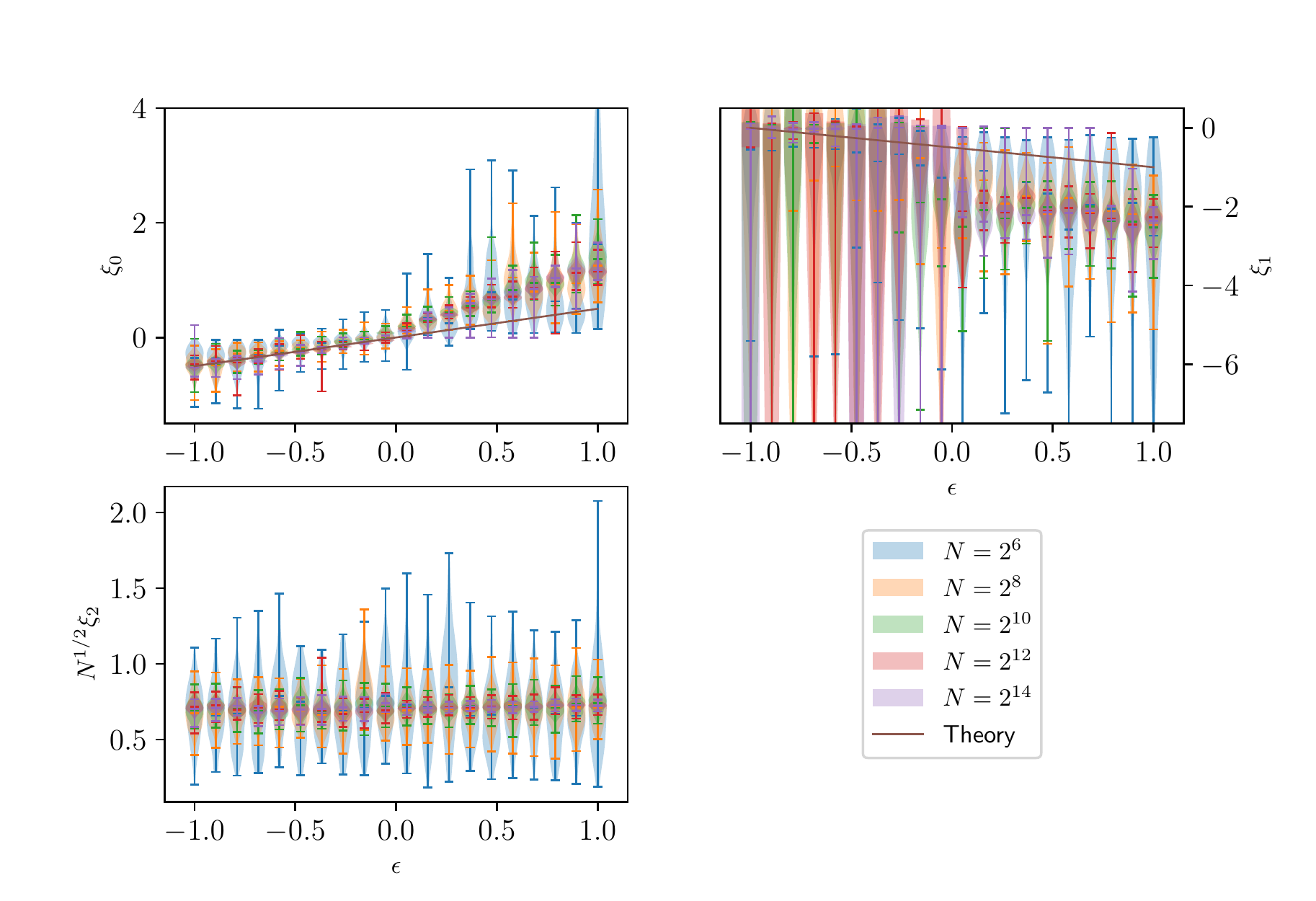}
    \vspace{-.3in}
    \caption{Overview of $\xi$ parameters inferred by Langevin regression from Kuramoto order parameter trajectories. As in Fig.~\ref{fig:r_avg_overviews}, for each value of $(N, \epsilon)$ we sample 100 Kuramoto systems, and for each system obtain an estimate of SDE parameters $\xi = (\xi_0, \xi_1, \xi_2)$. Vertical violins depict the distribution of the corresponding 100 quantities. ``Theory'' lines in the top panels are based on the Ott-Antonsen ansatz. $\xi_2$ is scaled by $N^{1/2}$ so that values for all $N$ are comparable, and we can see that $\xi_2$ is approximately consistent with $\xi_2 = \sigma N^{-1/2}$ for a global constant $\sigma\approx 1/\sqrt{2}$.
    }
    \label{fig:violin_plots_overview}
\end{figure}

An overview of parameter values inferred via Langevin regression based on trajectories of the Kuramoto order parameter are given in Fig.~\ref{fig:violin_plots_overview}. In the top two panels, the brown line depicts the value predicted by the theory derived by Ott-Antonsen, namely $\xi_0 = \epsilon/2$ and $\xi_1 = -(\epsilon+1)/2$. In the bottom-left panel, the horizontal line is at height $1/\sqrt{2}$.

The key result in Fig. \ref{fig:violin_plots_overview} is in the bottom-left panel, which depicts distributions of inferred values of the noise intensity, $N^{1/2}\xi_2$. Langevin regression clearly shows that a noise intensity scaling like $N^{-1/2}$ is consistent with the data. Moreover, the distributions of inferred values of $N^{1/2}\xi_2$ are centered on $1/\sqrt{2}$, for all $N$ and for all $\epsilon$. This remarkable regularity shows that when we regard the Kuramoto order parameter as following an SDE, its critical fluctuations can be explained by a simple (white) noise process applied to a drift term that undergoes a bifurcation.

On the subcritical side of the bifurcation, inferred values for $\xi_0$ cluster around the theoretical value, with a spread that shrinks as $N$ increases. For $\xi_1$, however, inferred values show much larger spread. This is because in the subcritical phase, $r$ is small, and so $r^3$ is very small. Thus its coefficient, $\xi_1$, is very weakly constrained and highly susceptible to blowup. The $\ell_1$ term in the cost function substantially mitigates this problem, but does not eliminate it entirely.

On the supercritical side, we also see noticeable trends with $\epsilon$ and $N$. Interestingly, inferred coefficient values do not cluster around the theoretical line. Rather, they are systematically farther from zero (i.e. $\xi_0$ is more positive and $\xi_1$ is more negative). 
This indicates that finite Kuramoto systems exhibit drift behavior that is nontrivially different from what is predicted in the $N\to\infty$ limit. Alternatively, it is possible that a different noise model is more appropriate, and that with such a noise model in place the drift coefficients would align with what we expect from Ott-Antonsen theory. In particular, it is possible that the fluctuations affecting the order parameter are correlated on a timescale long enough that white noise is not a good approximation, meaning that colored noise would be needed. Further research is warranted to derive possible higher-order corrections to the Ott-Antonsen theory.

\section{Discussion and Conclusions}

We have shown that it is possible to use Langevin regression to regress data from finite-size Kuramoto oscillator systems onto a simple class of SDEs consisting of linear and cubic drift terms and additive white noise. Since in the absence of noise, this model class contains the equation predicted by the Ott-Antonsen ansatz for the $N\to\infty$ limit, we can evaluate the limiting behavior of our learned models. We find that while the linear and drift terms exhibit the correct direction of dependence on $K$, there are substantial quantitative discrepancies even after accounting for variance across samples of natural frequencies.

Remarkably, however, we find that the fluctuations associated with finite system size are consistent with forcing by white noise at a magnitude that scales with $N$ as $N^{-1/2}$.
This demonstrates that finite-size effects in complex systems can be effectively modeled as stochastic forcing that follows CLT-like scaling, even as response fluctuations follow highly nontrivial scaling laws, especially near critical transitions.

\section*{Acknowledgements}
JS, SLB, and JNK acknowledge funding support from the Air Force Office of Scientific Research (AFOSR FA9550-19-1-0386).  JLC acknowledges funding support from the Department of Defense (DoD) through the National Defense Science \& Engineering Graduate (NDSEG) Fellowship Program.

\begin{appendices}
\section{Scaling of $\chi$ distributions}
\label{sec:chi-scaling-appendix}
We now discuss the scaling of the distributions of $\chi$ with $N$, for both large and small values of the bifurcation parameter, $\epsilon$. In particular, we see in Fig.~\ref{fig:r_avg_overviews} (right) that the distributions of $\chi$ are approximately constant with respect to $N$ when $\epsilon$ is large and negative, while they become more narrow with increasing $N$ when $\epsilon$ is large and positive.

\begin{figure}[hbt!]
    \centering
    \includegraphics{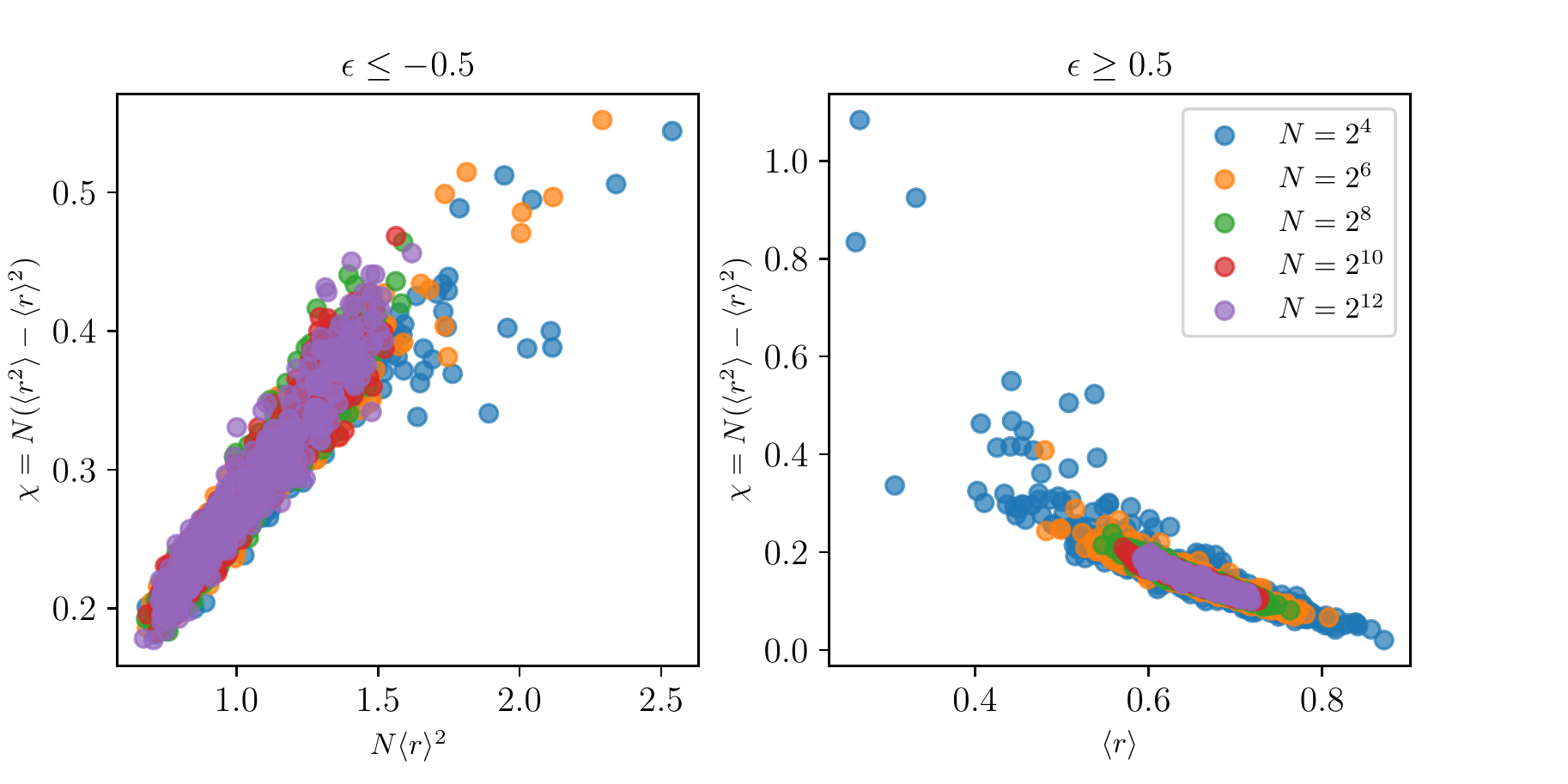}
    \caption{Relationship between $\langle r \rangle$ and $\chi$ in the subcritical (left, $\epsilon\le -0.5$) and supercritical (right, $\epsilon \ge 0.5$) regimes. Each dot corresponds a Kuramoto system, i.e. Eq.~(\ref{eq:kuramoto}) with a sample of natural frequencies $\{\omega_i\}$. In the subcritical regime, $\chi$ is positively correlated with $\langle r \rangle$, while the opposite is true in the supercritical regime.}
    \label{fig:sub-and-super-scaling-relationships}
\end{figure}

This phenomenon can be explained by the correlation between $\langle r \rangle$ and $\chi$ in the sub- and super-critical regimes, respectively, as visualized in Fig.~\ref{fig:sub-and-super-scaling-relationships}.

We can explain these relationships by appealing to approximate SDE models that hold when $r\approx 0$ and when $r\approx r_0 \neq 0$. We begin with the SDE~(\ref{eq:radial_SDE_form}):
\begin{equation}
    \du r = \left(\lambda r + \mu r^3 + {\sigma^2 \over 2r}\right) \du t + \sigma\du W
\end{equation}
First, when $r\approx 0$, as in the subcritical regime, the $r^3$ term is small and we omit it, to obtain
\begin{equation}
    \du r = \left(\lambda r + {\sigma^2 \over 2r}\right) \du t + \sigma\du W.
\end{equation}
In this case the steady-state PDF is
\begin{equation}
    p(r) = {-2\lambda \over \sigma^2} r \exp\left({\lambda r^2 \over \sigma^2}\right)
\end{equation}
and its first two moments are
\begin{align}
    \langle r \rangle &= {\sqrt{\pi}\over 2}{\sigma \over \sqrt{-\lambda}}\\
    \langle r^2 \rangle &= {-\sigma^2 \over \lambda} = {4 \over \pi} \langle r \rangle^2
\end{align}
which implies that $\chi = N(\langle r^2 \rangle - \langle r \rangle^2) = N({4 \over \pi}-1) \langle r \rangle^2$. In particular, the distribution of $\chi$ should scale with $N$ in the same manner as $N\langle r \rangle^2$. Since in the subcritical regime, correlations between phases are weak, it is reasonable to view $\langle r\rangle$ as the mean of $N$ i.i.d. samples, and therefore subject to the central limit theorem. The CLT would predict that $\langle r \rangle$ is approximately normally distributed with standard deviation $\mathcal{O}(1/\sqrt{N})$, so $N\langle r \rangle^2$ should be $\mathcal{O}(1)$ for all $N$.

In the supercritical regime, on the other hand, $r$ is not small. Rather, it falls in a narrow band around some nonzero value, $r_0$. In this case we suppose that the drift term of Eq.~(\ref{eq:radial_SDE_form}) can be approximated linearly:
\begin{equation}
    \du r = \overline{\lambda}(r-r_0) \du t + \sigma\du W.
\end{equation}
where $\overline{\lambda}$ depends on $\lambda, \mu, \sigma$, and $r_0$. In this case the steady-state PDF is Gaussian:
\begin{equation}
    p(r) = \sqrt{{\overline{\lambda} \over \pi \sigma^2}} \exp\left({\overline{\lambda}\over \sigma^2} (r-r_0)^2\right)
\end{equation}
which (neglecting the bounds $0<r<1$) has moments
\begin{align}
    \langle r \rangle &= r_0\\
    \langle r^2 \rangle &= {-\sigma^2 \over 2\overline{\lambda}} + r_0^2
\end{align}
and thus $\chi = -N\sigma^2 / \overline{\lambda}$.

Next, we can examine the dependence of $\chi$ on $r_0$ by partial differentiation:
\begin{align}
    {\partial \chi \over \partial r_0} &= {\partial \chi \over \partial \overline{\lambda}}{\partial \overline{\lambda} \over \partial r_0}\\
    &= {N\sigma^2 \over 2\overline{\lambda}^2} \left[ 6\mu r_0 + {\sigma^2 \over r_0^3}\right].
\end{align}
Clearly $N\sigma^2 / 2\overline{\lambda}^2$ is positive; we now show that the term in square brackets is negative. To do this we multiply it by $r_0^3$ and obtain
\begin{align}
    r_0^3 \left[ 6\mu r_0 + {\sigma^2 \over r_0^3}\right] &= 6\mu r_0^4 + \sigma^2\\
    &= 6\left({-\sigma^2 \over 2} - \lambda r_0^2 \right) + \sigma^2\\
    &= -2\sigma^2 - 6\lambda r_0^2 < 0
\end{align}
where we have used the condition that $r_0$ is a root of the drift term, namely that $\lambda r_0^2 + \mu r_0^4 + \sigma^2/2 = 0$, and that $\lambda>0$ since we consider the supercritical regime. Thus, $\partial \chi / \partial r_0 <0$, and indeed we see that $\chi$ and $\langle r \rangle$ are negatively related in the supercritical phase. Thus we should expect the distribution of $\chi$ to be comparable to the distribution of $\langle r \rangle$, i.e. it should become narrower with larger $N$.

These results can be understood in terms of the population of oscillators. In the subcritical regime, when coupling is weak, both $\langle r\rangle$ and $\chi$ are driven by the amount of correlation among the oscillators' phases, and are therefore positively related. In the supercritical regime, on the other hand, fluctuations are driven by the drifting subpopulation of oscillators, i.e. those whose frequency does not lock to the global frequency. The larger the value of $r$, the fewer oscillators remain drifting, and therefore the fewer oscillators there are to contribute to fluctuations in $r$.

\section{Normally distributed natural frequencies}
Here we consider stochastic modeling of the order parameter for Kuramoto systems where the natural frequencies are drawn not from a Cauchy distribution, but a normal (Gaussian) distribution. For comparability with the Cauchy-distributed case, we take the standard deviation $\sigma$ to be $\sqrt{\pi/8}$, so that the critical coupling strength is $K_c = 1$. As before, we take 20 values of $\epsilon=K-K_c$ evenly spaced between $-1$ and $1$, and for each value of $\epsilon$ perform 100 replicate experiments consisting of sampling natural frequencies and initial conditions, and integrating the equations of motion. We do this for the number of oscillators $N = \{64, 256, 1024, 4096\}$.

\begin{figure}
    \centering
    \includegraphics[width=\textwidth]{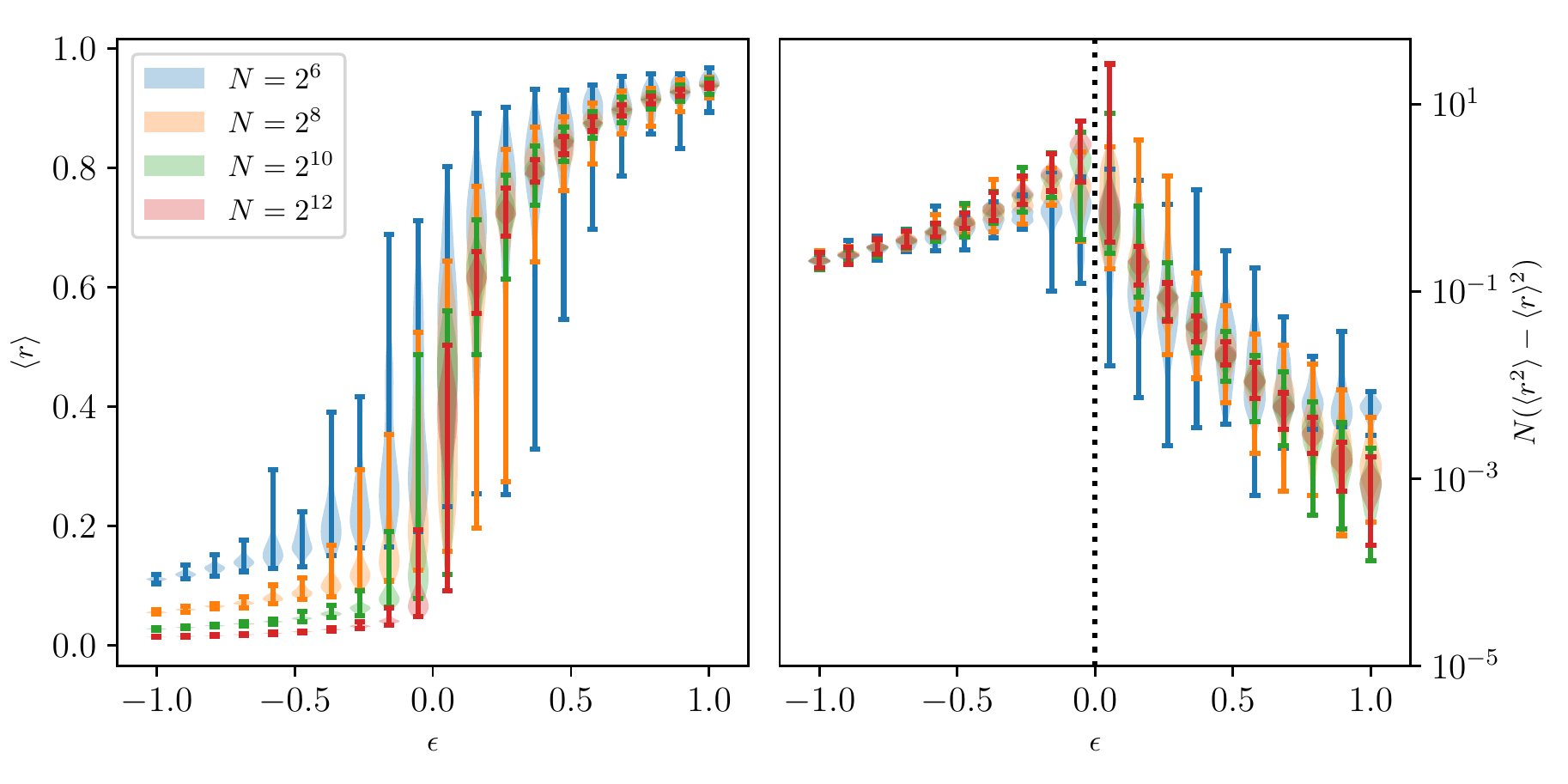}
    \caption{Distributions of the temporal average $\langle r \rangle$ and scaled temporal variance $\chi = N(\langle r^2\rangle - \langle r \rangle ^2)$ of the order parameter as a function of number of oscillators $N$ and bifurcation parameter $\epsilon$. Trends agree qualitatively with those for the Cauchy case depicted in Fig.~\ref{fig:r_avg_overviews}, including the width of the distribution of $\chi$ above vs. below the bifurcation. Note that for small $N$ and large $\epsilon$, many sampled systems achieve complete synchronization, leading to nearly-constant $r(t)$ and vanishingly small $\chi$. We omit these from the statistics here, and therefore the remaining cases have comparatively large values of $\chi$ (see the $N=2^6$ cases for $\epsilon\approx 1$).}
    \label{fig:r_and_chi_overview_normal}
\end{figure}

In Fig.~\ref{fig:r_and_chi_overview_normal} we depict distributions of the temporal mean $\langle r \rangle$ and temporal variance $\langle (r-\langle r\rangle)^2\rangle$ of the order parameter as a function of $N$ and $\epsilon$ (cf. Fig.~\ref{fig:r_avg_overviews} for the Cauchy-distributed case). As there is no known closed-form expression for the value of the order parameter as $N\to\infty$ in the case that natural frequencies are normally distributed, we omit the $N\to\infty$ theory line in the left panel. Note that the value of $\chi$ decays with $\epsilon$ more quickly than it does in the case of Cauchy-distributed natural frequencies. This is because when natural frequencies are normally distributed, much less coupling is required to synchronize a given fraction of the population of oscillators than when natural frequencies are Cauchy-distributed.

\begin{figure}
    \centering
    \includegraphics[width=\textwidth]{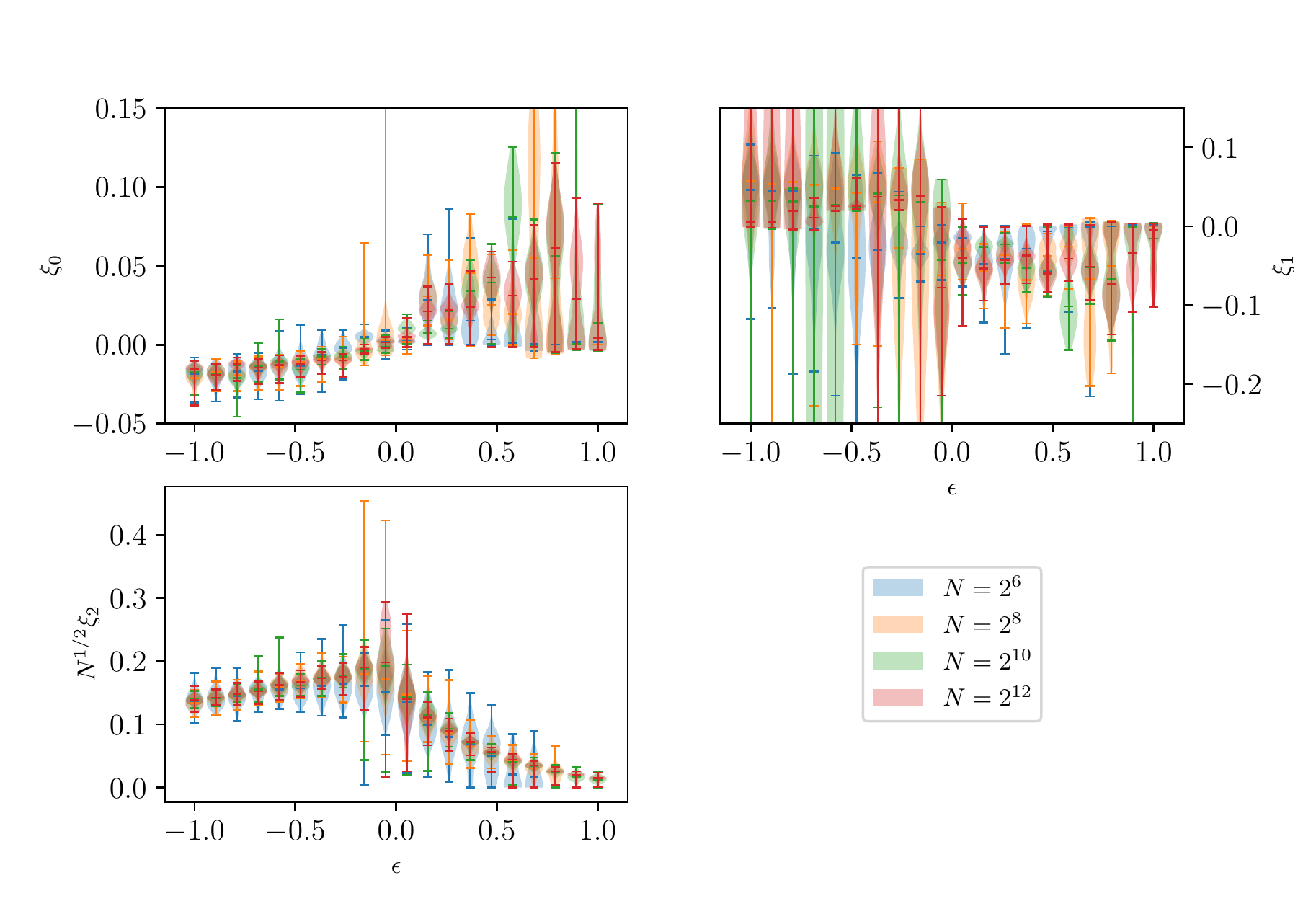}
    \caption{Distributions of inferred drift and diffusion coefficients for Kuramoto systems with normally-distributed natural frequencies (cf. Fig.~\ref{fig:violin_plots_overview}).}
    \label{fig:normal_xi_overview}
\end{figure}

In Fig.~\ref{fig:normal_xi_overview} we show the distribution of inferred drift and diffusion coefficients for Kuramoto systems with normally distributed natural frequencies, obtained via Langevin regression. Several key differences from the Cauchy-distributed case (Fig.~\ref{fig:violin_plots_overview}) are apparent. First, the absolute size of the inferred coefficients is smaller (compare vertical scales). Next, the distributions of the (rescaled) noise strength $N^{1/2}\xi_2$ are not constant with $\epsilon$. Rather, their centers show mild increase with $\epsilon$ up to the critical point, and decay towards zero as $\epsilon$ grows above the critical point. This indicates that when natural frequencies are normally-distributed, the effective strength of noise decreases as more of the population becomes synchronized. Perhaps surprisingly, this is not the case when natural frequencies are Cauchy-distributed, owing to the extreme heterogeneity of the Cauchy distribution.

Similarly to the Cauchy-distributed case, we also see significant noise in the recovery of the cubic coefficient in the subcritical regime. On the other hand, we also observe irregularity at the far-supercritical side of the bifurcation. This is due to the fact that when $\epsilon$ is large, the system can easily become completely synchronized, and the order parameter becomes constant with time. In these cases a stochastic model is clearly inappropriate and the inferred coefficients lose their meaning.

\end{appendices}

\bibliographystyle{plain}
 \begin{spacing}{.9}
 \small{
 \setlength{\bibsep}{6.5pt}
 \bibliography{references,references2}
 }
 \end{spacing}
\end{document}